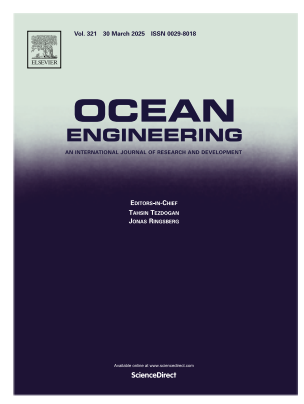

Research paper

# A distributed motion planning approach to cooperative underwater acoustic source tracking

Andrea Tiranti [a,*], Francesco Wanderlingh [a,b], Enrico Simetti [a,b], Marco Baglietto [a,b], Giovanni Indiveri [a,b], Antonio Pascoal [c]

[a] DIBRIS, University of Genova, Via all'Opera Pia 13, 16145 Genova, Italy, Genova, 16145, Italy
[b] ISME, Interuniversity Center of Integrated Systems for the Marine Environment, Via all'Opera Pia 13, 16145 Genova, Italy, Genova, 16145, Italy
[c] ISR, Institute for Systems and Robotics, Instituto Superior Técnico, Av. Rovisco Pais 1, Lisbon, 1049-001, Portugal

ARTICLE INFO

ABSTRACT

Accurate tracking of underwater acoustic sources is critical for a variety of marine applications, yet remains a challenging task due to communication constraints and environmental uncertainties. In this regard, this paper addresses the problem of underwater acoustic source tracking using a team of autonomous underwater vehicles (AUVs). The core idea is to optimize the guidance of each agent to achieve coordinated motion planning that leads to optimal geometric configurations with respect to the target, thereby enhancing tracking performance. To tackle this, we propose a Distributed Model Predictive Control (DMPC) framework to improve performance and robustness. The control problem is formulated as a multi-objective optimization task, incorporating geometric observability, proximity to the target, and communication connectivity. A Receding Horizon Control (RHC) approach, coupled with an Unscented Transform (UT)-based prediction scheme, is employed to ensure long-term tracking accuracy while accounting for uncertainties. The optimization is distributed using the sequential multi-agent decision-making framework, combined with the Time-Division Multiple Access (TDMA) communication protocol. The proposed methodology is implemented in a simulation environment that accounts for the constraints of acoustic communication. The approach is compared with existing methods such as decentralized MPC and Particle Swarm Optimization (PSO).

## 1. Introduction

In recent decades, there has been a growing imperative to monitor marine areas in the context of various application domains, including biodiversity research and conservation (Mooney et al., 2020), and monitoring of civilian activities (Baumgartner et al., 2018). In such scenarios, a current open problem is tracking acoustic sources, historically addressed with underwater sensor networks equipped with hydrophones. The advent of Autonomous Underwater Vehicles (AUVs) represents a fundamental step for acoustic monitoring systems since they afford end-users a completely new level of autonomy and drastically increase performance. Therefore, recently the effort of the scientific community has moved toward the design of Underwater Mobile Sensor Networks (UWMSN)(Ferri et al., 2017). Fig. 1 gives a qualitative representation of an operational scenario in which UWMSNs can be beneficial. In this work, we control a group of (AUVs) agents in which the latter play the roles of moving nodes of the network. Since Acoustic Vector Sensors (AVS) and Streamers (arrays of hydrophones) represent the current most prominent technologies in acoustic sensing (Cao et al., 2017), we assume that each agent measures the Direction of Arrival (DoA) of the acoustic signal emitted by the target. This assumption is valid for any Passive Acoustic Monitoring (PAM) (Wolek et al., 2019) application where the target is an underwater vehicle, a mammal or a tagged marine animal (Klinck et al., 2016). Estimating the dynamic of the target with bearing-only information is by itself a challenging task since the observability of the target is not guaranteed unless the sensor *outmaneuvers* the target by performing sufficiently exciting trajectories (Farina, 1999). To mitigate this, multiple sensors are often employed, and the gathered information is fused to estimate the target's state (Han et al., 2019).

Building on this, the problem of distributed tracking has been widely explored, with two main strategies for information fusion: *share and estimate* and *estimate and share* approaches (Allotta et al., 2021) . In the latter, each agent independently estimates the target's state using local measurements and then exchanges these estimates for refinement via weighting strategies (Radtke et al., 2019; Nguyen et al., 2021). This approach, however, requires agents to follow exciting trajectories to ensure observability, which can be an energy- and time-intensive demand, particularly problematic in passive acoustic monitoring (PAM) scenarios (Shinzaki et al., 2013). Additionally, underwater measurements are inherently noisy and prone to outliers (Borker et al., 2015), making it advantageous to gather data from diverse positions. Given these constraints, we adopt a *share and estimate* approach, where raw measurements are exchanged among agents. In this setting, the spatial configuration between sensors and the target becomes crucial for observability.

* Corresponding author.
*E-mail addresses:* andrea.tiranti@edu.unige.it (A. Tiranti), francesco.wanderlingh@unige.it (F. Wanderlingh), enrico.simetti@unige.it (E. Simetti), marco.baglietto@unige.it (M. Baglietto), giovanni.indiveri@unige.it (G. Indiveri), antonio.pascoal@tecnico.ulisboa.pt (A. Pascoal).

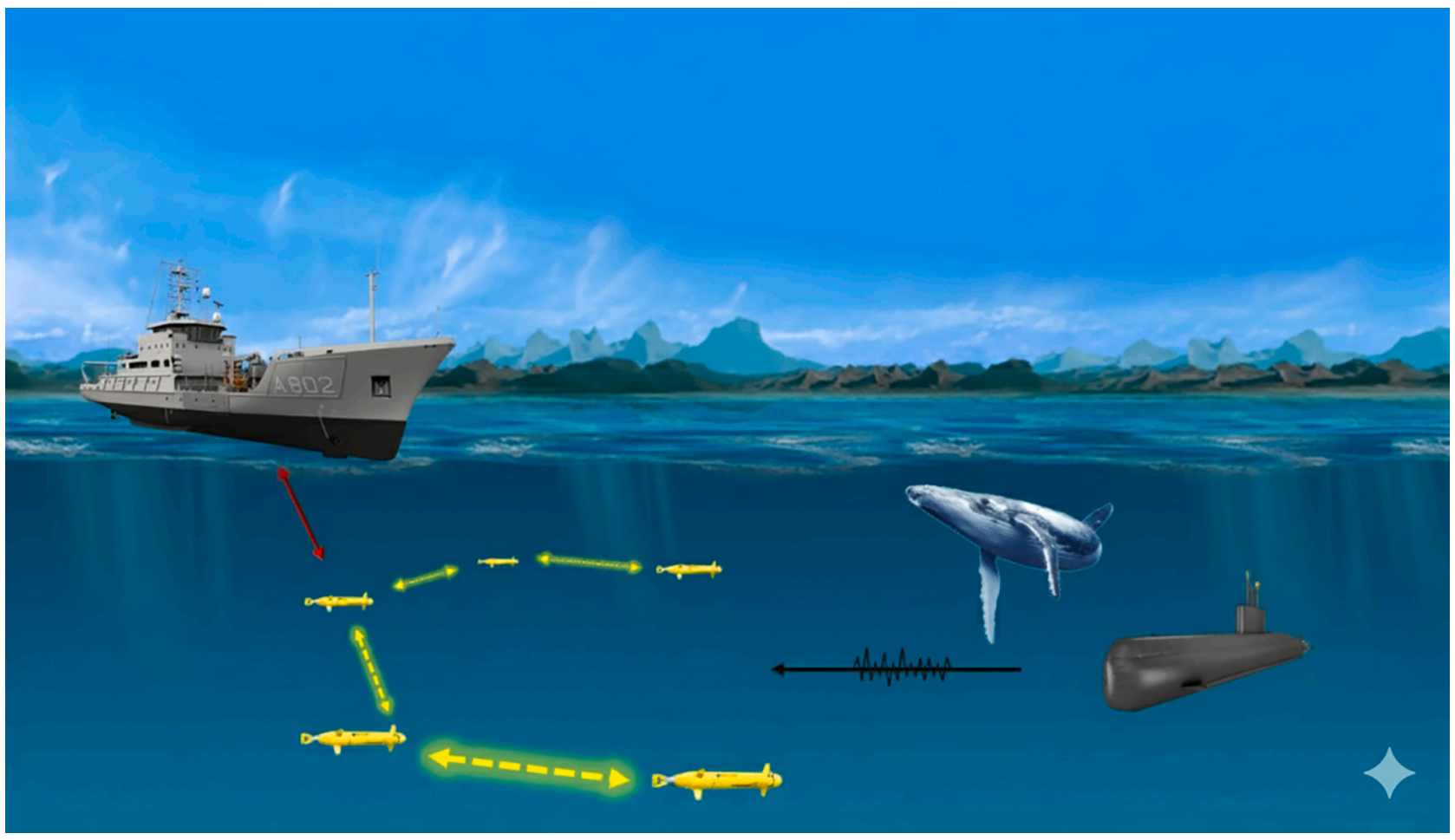

**Fig. 1. Acoustic Target Monitoring using an Underwater Mobile Sensor Network (UMSN):** The image shows a qualitative representation of an artist's rendition of a Passive Acoustic Monitoring (PAM) scenario using a team of Autonomous Underwater Vehicles (AUVs). The latters are employed as moving nodes of the network and equipped with Acoustic Vector Sensors (AVS) that can measure the direction of a detected acoustic signal.

Indeed, as shown in Dogancay (2022), the geometry between sensors and the target directly impacts observability and the performance of distributed estimation algorithms, due to its effect on the *conditioning* of the estimation problem. A well-conditioned estimation problem is less sensitive to noisy or corrupted data, as indicated by a low condition number (Tiranti et al., 2023). The Fisher Information Matrix (FIM) further supports this analysis by linking observability to the agents' relative motion and range-based information (Moreno-Salinas et al., 2016). It has been demonstrated that an optimal geometric configuration not only enhances estimation accuracy but also improves the probability of detecting an acoustic source (Ferri et al., 2020). Despite its importance, the pursuit of optimal geometry is often neglected, especially in scenarios where agents prioritize reaching the target over maintaining observability (Kim, 2019). Common approaches assume fixed formations to preserve a minimum baseline (Zolich et al., 2017; Li et al., 2021a), which limits maneuverability and adaptability. Dynamic formation updates are rarely addressed, with most works, such as Zhu et al. (2022), relying only on discrete formation switching.

To overcome these limitations, more recently, the scientific community has moved toward implementing more sophisticated control strategies to address the guidance and control problem in underwater cooperative target tracking. This challenge has been investigated across disciplines, such as optimal sampling in oceanographic applications (Alvarez and Mourre, 2012) and sensor management (Hero and Cochran, 2011), where the control of sensor degrees of freedom is framed as a stochastic control problem aiming to optimize mission-specific cost functions. The objective is to determine optimal or suboptimal policies that guide sensor configuration using prior measurements and environmental models, while adhering to operational constraints. However, real-world implementation remains challenging due to unreliable and intermittent acoustic communication (Yan et al., 2018), which disrupts information flow across agents and results in partial or inconsistent environmental knowledge. Graph theory provides a valuable framework for modeling sensor networks and ensuring connectivity, even in intermittently connected graphs, by satisfying constraints like *connection on average* (Moreau, 2005) or the dynamic graph assumptions in Riehl et al. (2011). These principles support the convergence of local knowledge across agents within distributed systems, making them particularly relevant for mobile sensor networks. Altogether, these considerations suggest framing the cooperative tracking problem as a Multi-Objective Optimization (MOO) problem.

Regardless of the specific cost functions, a key factor in such optimization strategies is the planning horizon, particularly in guidance problems where long-term objectives must be considered. Actions that may not immediately reduce the costs can still be valuable over an extended horizon. While theoretical approaches such as Markov Decision Processes (MDPs) (Li et al., 2021b) and dynamic programming (Bertsekas and White, 1977) offer optimal solutions, the complexities of the underwater domain, including disturbances and partial observability, make Partially Observable MDPs (POMDPs) (Lauri et al., 2022) a suitable yet computationally demanding framework. For resource-limited robotic agents, approximate methods like rollout algorithms or Model Predictive Control (MPC) (Bertsekas, 2012) offer a practical alternative. In particular, receding horizon control schemes strike a balance by computing optimal action sequences over a moving window. The length of the planning horizon is a trade-off between the desire to explore the decision space (favoring longer horizons) and the limitations of feedback based on uncertain estimates (favoring shorter ones).

Finally, beyond the estimation and control aspects, the implementation must also account for system-level considerations. To avoid the *single-point of failure* issue, we adopted a distributed implementation, as maintaining connectivity to a central node is often impractical in underwater environments. While early solutions to optimization problems in sensor networks used centralized approaches, particularly for formation-based missions, a centralized scheme, like the Deep Reinforcement Learning-based POMDP solution in Song et al. (2022), risks reducing system endurance and is less suitable for anchor-free scenarios (Hu et al., 2021). Distributed approaches offer greater resilience, scalability, and efficiency by sharing network resources (Charalambous and Ahmed, 2017). However, distributed optimization in the joint action domain remains challenging due to exponential growth in complexity with the number of agents under naïve *greedy* strategies. To address this, we rely on *sequential multi-agent decision making* to design our distributed control framework (Gmytrasiewicz and Doshi, 2005). Using this approach, we can distribute the computational load as exemplified in Li et al. (2022), where a cooperative tracking problem is modeled through POMDPs and solved using sequential framework. We approached cooperative passive acoustic target tracking as a POMDP solved via MPC. While Multi-Agent Reinforcement Learning (MARL) is gaining popularity for similar problems, its application in underwater environments faces major challenges, including the need for large interaction datasets and the difficulty of sim-to-real transfer (Tong et al., 2023). Unlike many MARL studies that assume ideal communication (Yang et al., 2021; Zhu et al., 2024a), our system involves packet loss, delays, and partial observability, conditions under which MARL struggles to have reliable performance (Zhang et al., 2024).

### 1.1. Contributions

Bearing the above references in mind, the contribution of this paper is a Distributed Model Predictive Control (DMPC) approach for multiple-vehicle motion planning in underwater cooperative target tracking scenarios. The DMPC is used as a planning module within the distributed guidance, navigation, and control (GNC) architecture. The objective is to act on the guidance of the vehicles to increase the robustness and performance of a UWMSN in acoustic target tracking. Three strategies are used to optimize the motion of the acoustic nodes: (i) the agents must improve their geometry with respect to the target position trying to maintain an optimal configuration with respect to a certain cost criterion, (ii) the agents must reduce their distance to remain in a certain vicinity of the target, (iii) the agents must maintain or improve the connectivity of the graph to reduce the probability of losing communication. An important assumption adopted in this work is that the agents access the acoustic communication channel following the contention-free Time-Division Multiple Access (TDMA) protocol.

To implement the proposed stochastic optimization, we designed an MPC scheme since we are interested in the long-term tracking performance. To make the optimization problem tractable, we use the theory of Unscented Transform (UT) to predict the state of the system involved during the lookahead optimization. To solve the MMO in a distributed way, the DMPC is implemented in a sequential multi-agent planning framework compliant with the TDMA communication protocol. Considering the state of the art presented in Section 1, the specific contributions of this manuscript can be summed up as follows:

- The proposed approach integrates a *perception module* based on a distributed tracking filter that uses the *share and estimate* paradigm to address the bearing-only estimation problem. Such a methodology reduces the impact of noisy measurements and outliers, with the advantage of obtaining measurements from different positions, thus allowing for observability without pursuing exciting trajectories, unlike Shinzaki et al. (2013), Song et al. (2022), Crasta et al. (2018).
- In contrast with the work in Zhu et al. (2024b), Cao and Guo (2019), Kim (2019), the proposal is a fully distributed solution based on a sequential multi-agent control framework. Basically, the proposed framework extends the data-driven MPC strategy in Ferri et al. (2018), which directly controls the heading of a single AUV in a bistatic sonar setup. Here, the same principle is generalized to a distributed, passive sonar context, where data exchanged among agents guides a high-level MPC planner that generates optimal motion trajectories subsequently tracked by low-level controllers.
- Unlike the following works Nguyen et al. (2021), Fan et al. (2024), and Li et al. (2020), we designed the methodology to be robust to latencies and packet losses. We tested it in a custom simulation environment that explicitly accounts for the major limitations associated with acoustic communication.

The remainder of the paper is organized as follows. Section 2 outlines the adopted methodology, and Section 3 presents and discusses the simulation results that support the proposed approach. Lastly, conclusions and future work are addressed

## 2. Methodology

### 2.1. Symbols and notations

In this section, we define the symbols and notations used throughout the manuscript.

### 2.2. Preliminaries

In cooperative target tracking scenarios, the goal is to estimate the state (position and velocity) of an underwater moving acoustic source. Let $\xi = (\boldsymbol{p}^\top, \boldsymbol{v}^\top)^\top \in \mathbb{R}^4$ represent the state of the target moving in 2D, where $\boldsymbol{p} = [x, y]^\top$ denotes the inertial position and $\boldsymbol{v} = \dot{\boldsymbol{p}}$ denotes the inertial velocity. For the time being, consider a single underactuated AUV acting as a tracker (equipped with an acoustic sensor) with the state described by $\boldsymbol{s} = [\boldsymbol{p}_s^\top, \theta]^\top$, where $\boldsymbol{p}_s = [x_s, y_s]^\top$ and $\theta$ are the vehicle´s position and heading respectively. At a kinematic level, the input of the moving agent is given by $\boldsymbol{u} = [u, r]^\top$, where $u$ and $r$ are the surge speed and yaw rate, respectively. Notice that we are adopting a simplified kinematic model for the AUV where *sway* speed is not considered. The kinematics of the sensor are thus given by

$$\begin{bmatrix} \dot{x}_s \\ \dot{y}_s \\ \dot{\theta}_s \end{bmatrix} = \boldsymbol{\Theta}(\theta_s) \begin{bmatrix} u \\ 0 \\ r \end{bmatrix} \tag{1}$$

where $\boldsymbol{\Theta}(\theta_s) \in \mathbb{R}^{2\times 2}$ is an orthonormal rotation matrix that converts coordinates from the vehicle frame to the reference inertial frame. The overall problem is sketched in Fig. 2. Since we can only measure the DoA of the acoustic signal from the target (given in terms of bearing angle $\beta$), it is useful to define the relative angle between the target and the vehicle as

$$\beta(t) = \operatorname{atan2}(p_y(t) - p_{s,y}(t), p_x(t) - p_{s,x}(t)) \; . \tag{2}$$

The target's dynamics, assuming it moves with constant velocity, are described as

$$\dot{\xi}(t) = \boldsymbol{A}\xi(t) = \begin{bmatrix} \boldsymbol{O} & \mathbb{I}_2 \\ \boldsymbol{O} & \boldsymbol{O} \end{bmatrix} \xi(t) \tag{3}$$

with output equation

$$h(t) = \beta(t) + \nu(t), \tag{4}$$

where $\nu(t)$ is the measurement noise that, given the properties of the PAM sensors, is characterized by a uniform distribution with bounds $\pm\sigma_\nu$. Eq. (4) is a non-linear function of the target state vector $\xi(t)$, thereby making the overall system a nonlinear state-space model. Notice the bearing measurement in (4) can be rewritten as

$$\frac{\sin(h(t) - \nu(t))}{\cos(h(t) - \nu(t))} \triangleq \frac{sin(\beta(t))}{cos(\beta(t))} \triangleq \frac{p_y(t) - p_{s,y}(t)}{p_x(t) - p_{s,x}(t)} \; . \tag{5}$$

After some algebraic manipulations, as shown in Aidala (1979), the output equation may be rewritten as

$$\boldsymbol{\eta}^\top \boldsymbol{p}_s(t) = \boldsymbol{\eta}\boldsymbol{H}\,\boldsymbol{x}(t) + f(\xi(t), \nu(t)) \; , \tag{6}$$

where

$$\boldsymbol{\eta} = \begin{bmatrix} \sin\beta(t) \\ \cos\beta(t) \end{bmatrix}, H = \begin{bmatrix} 1 & 0 & 0 & 0 \\ 0 & 1 & 0 & 0 \end{bmatrix} \tag{7}$$

Letting $z(t) = \boldsymbol{\eta}^\top \boldsymbol{p}_s(t)$ and $\boldsymbol{C} = \boldsymbol{\eta}^\top \boldsymbol{H}$ we obtain an output model of the form

$$z(t) = \boldsymbol{C}(\boldsymbol{\xi}(t))\boldsymbol{\xi}(t) + f(\boldsymbol{\xi}(t), \upsilon(t), \boldsymbol{s}(t)) \tag{8}$$

with output matrix

$$\boldsymbol{C}(\boldsymbol{\xi}(t)) = \begin{bmatrix} \sin\beta(t) & -\cos\beta(t) & 0 & 0 \end{bmatrix} \tag{9}$$

and noise (Lingren and Gong, 1978):

$$f(\boldsymbol{\xi}(t), \upsilon(t), \boldsymbol{s}(t)) = \left\| \boldsymbol{p}(t) - \boldsymbol{p}_s(t) \right\| \nu(t) \; . \tag{10}$$

| | |
|---|---|
| $x$ | Scalar variable |
| $\mathbf{x}$ | Vector variable |
| $\bar{X}$ | Vector of vectors (e.g., representing a set of vectors or time-varying vector sequences) |
| $\mathbf{X}$ | Matrix (bold capital letters represent matrices) |
| $\mathcal{X}$ | Set (uppercase letters in italic represent sets) |
| $\mathbf{I}$ | Identity matrix |
| $\mathbf{P}$ | Covariance matrix |
| $\mathbb{R}^n$ | Euclidean space of dimension $n$ |
| $\|\cdot\|$ | Euclidean 2-norm of a vector $\mathbf{x}$ |
| $\hat{}$ | Estimated/expected quantity |
| $^*$ | Desired/optimal quantity |

The measurement noise and the output matrix are now state-dependent. The distributed estimation algorithm uses the *share and estimate* paradigm that foresees the transmission of measurements between agents. It follows that each agent collects a set of $M$ measurements composed of local and received measurements

$$z = [z(t_0)\ z(t_1)\ \dots\ z(t_M)]^\top \in \mathbb{R}^M, \tag{11}$$

that can be processed to solve a least square estimation problem. Existing literature indicates that outliers and non-uniform time sampling can destabilize estimation algorithms such as the Extended Kalmen Filter (EKF) (Sinopoli et al., 2004). To this end, we used the theory of batch-recursive least square estimation presented in Di Lillo et al. (2024). We avoid going into detail on this topic for obvious reasons of space, since the focus of this work is motion planning. For the sake of clarity, the key concepts will be presented. The main idea of the batch-recursive estimator is to process multiple measurements together considering a moving time window to reduce the impact of noise and outliers. Given the set of measurements $z$, is possible to compute an associated least square regressor $\mathbf{\Phi} \in \mathbb{R}^{P\times 4}$. The estimate of the target state is obtained by solving († stands for pseudoinverse)

$$\hat{\xi}(t) = e^{\mathbf{A}(t-t_0)}\mathbf{\Phi}^\dagger z, \tag{12}$$

where $t_0$ is the timestamp of the oldest measure in the regressor. The proposed method is valid when the assumption of constant velocity holds in the estimation moving time window. The rationale behind this choice is that in acoustic tracking scenarios, it is difficult to formulate specific dynamic models for the target.

Finally, when an agent collects enough measurements, an estimate $\hat{\xi}$ is obtained by first computing the state in the past instant $t_0$ as

$$\hat{\xi}(t_0) = \mathbf{\Phi}^\dagger z, \tag{13}$$

and then propagating it forward in time using the estimation of the velocity

$$\hat{\xi}(t) = e^{\mathbf{A}(t-t_0)}\hat{\xi}(t_0)\ . \tag{14}$$

*Remark:* Given that the output uncertainty $v$ is not guaranteed to be normally distributed, the adopted Weighted Least Squares (WLS) estimator (with weight matrix $R^{-1}$ equal to the inverse of the output uncertainty covariance) is not granted to be optimal, but eventually it is only BLUE (Best Linear Unbiased Estimator) if $v$ has zero mean. The covariance associated with the estimation is

$$\mathbf{P} = (\mathbf{\Phi}^\top \mathbf{R}^{-1}\mathbf{\Phi})^{-1} \mid \mathbf{R} = diag(\sigma_v) \in \mathbb{R}^{M\times M}. \tag{15}$$

### 2.3. Extension to the multi-trackers case

Since we will be dealing with a multi-agent sensor system, it is useful to consider a corresponding graph that represents the underlying communication topology. To this end, we define an *undirected graph* $\mathcal{G} = \mathcal{G}(\mathcal{V}, \mathcal{E})$, where $\mathcal{V} = \{1, \dots, n\}$ represents the set of vertices, $\mathcal{E} \subseteq \mathcal{V} \times \mathcal{V}$ denotes the set of edges, and $n$ corresponds to the number of nodes

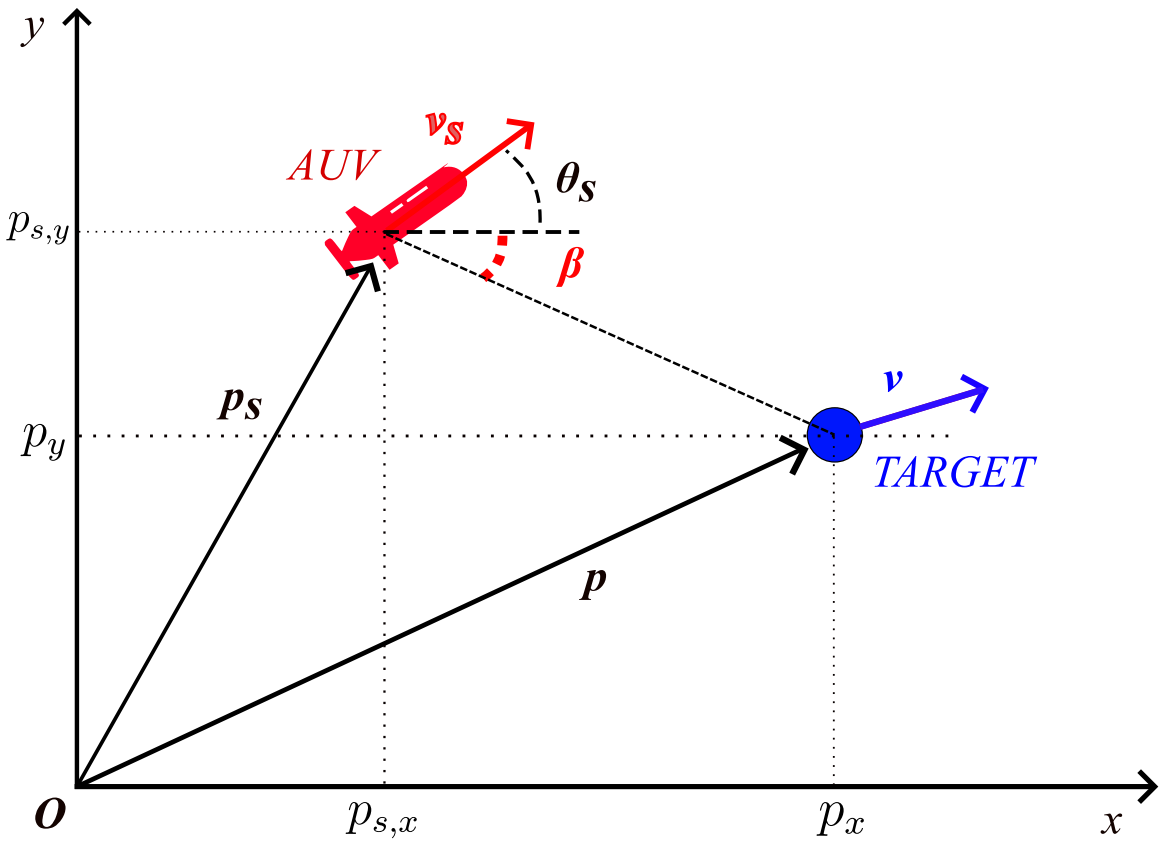

**Fig. 2. Single tracker estimation problem:** Sketch of the Bearing-Only Tracking (BOT) problem with one tracker: the blue dotted line represents the bearing angle, and the grey dotted line represents the Line of Sight (LOS) between the agent and the target.

(agents) in the network. We can define $\mathbf{\Lambda}$ as the adjacency matrix of $\mathcal{G}$ such that

$$\mathbf{\Lambda} = \left[a_{ij}\right] \in \mathbb{R}^{n\times n}, a_{ij} = \begin{cases} 1 & (j,i) \in \mathcal{E} \\ 0 & otherwise \end{cases} \tag{16}$$

Let us define the set of neighbours of agent $i$ as

$$\mathcal{N}_i = \{j \in \mathcal{V} : (j,i) \in \mathcal{E} \wedge (i,j) \in \mathcal{E}\}. \tag{17}$$

The degree matrix of the graph is defined as

$$\mathbf{\Delta} = \left[\delta_{i,j}\right], \delta_{i,j} = \begin{cases} |\mathcal{N}_i| & i = j \\ 0 & otherwise \end{cases}, \forall i = [1,n] \tag{18}$$

where the symbol $|\cdot|$ represents the cardinality operator. The (weighted) graph Laplacian matrix associated with $\mathcal{G}$ is defined as

$$\mathbf{L}(\mathcal{G}) = \mathbf{\Delta} - \mathbf{\Lambda}, \tag{19}$$

The goal during *cooperative target tracking* is to reach a consensus about the target state such that

$$\lim_{t\to+\infty} \left\|\hat{\xi}_i(t) - \xi(t)\right\| < \epsilon, \forall i \in [1,n] \quad s.t. \quad \epsilon > 0, \tag{20}$$

where $\hat{\xi}_i(t)$ is the estimation of the target state done by agent $i$. Using *graph theory* we can formulate a distributed form of closed-loop consensus dynamics as

$$\dot{\hat{\xi}}_i(t) = \mathbf{A}\mathbf{\Phi}_i^\dagger z_i + \mathbf{\Omega}_i \sum_{j\in N_i(t)} a_{i,j}(\hat{\xi}_i(t) - \hat{\xi}_j(t)), \tag{21}$$

where $z_i$ is the set of measurements available to agent $i$ as in (11), with associated regressor $\mathbf{\Phi}_i$, where $\mathbf{\Omega}_i \in \mathbb{R}^{4\times 4}$ is a consensus gain matrix.

In our case study, the connectivity of a graph depends on the Signal-to-Noise ratio (SNR) between nodes of the graph. Assuming a fixed transmission frequency, we can compute the SNR between two generic nodes

of the networks as

$$\rho_{ij} = SL - TL - NL + DI, \tag{22}$$

where $SL$ is the source level, $NL$ is the noise level, $DI$ is the directivity index related to the acoustic modem, and $TL$ is the transmission loss, which, according to Stojanovic and Preisig (2009), can be computed empirically as

$$TL = 20\log(d_{ij}) + d_{ij} \times \alpha(f) \times 10^{-3}, \tag{23}$$

where

$$\alpha(f) = 0.11\frac{f^2}{1+f^2} + 44\frac{f^2}{4100+f^2} + 2.75\times 10^{-4} f^2 + 0.003, \tag{24}$$

where $f$ is the acoustic center frequency in kHz of the transducer of the acoustic modem. The elements $l_{ij}$ of the Laplacian matrix of the graph in (19) can be rewritten as

$$l_{ij} = \begin{cases} -\rho_{ij} & i \neq j, \rho_{ij} \geq DT \\ \sum_{k=1}^{k=n} \rho_{ik} & i = j, \rho_{ij} \geq DT \\ 0 & i \neq j, \rho_{ij} < DT \end{cases} \tag{25}$$

where $DT$ is a detection threshold for the acoustic modem. The second smallest eigenvalue of the Laplacian matrix, often called the algebraic connectivity or Fiedler value, yields information about the connectivity of the graph. A larger $\sigma_2(\boldsymbol{L}(\mathcal{G}))$ indicates a stronger connectivity. If $\sigma_2(\boldsymbol{L}(\mathcal{G})) = 0$, the graph is disconnected.

To have a bounded eigenvalue, it is convenient to define an ideal value $\rho_M$ for the SNR to normalize the elements $l_{ij}$, i.e., when the transmission loss is zero $TL = 0$, or some desired values. The Laplacian matrix of the graph now becomes

$$\boldsymbol{L}(\mathcal{G}) = \left[\frac{l_{ij}}{\rho_M}\right] \in \mathbb{R}^{n\times n}. \tag{26}$$

### 2.4. The distributed model predictive control (DMPC) scheme

As described in (1), the state of agent $i$ is represented as $\boldsymbol{s}^i = [\boldsymbol{p}_s^i, \theta_s^i]^\top \in \mathcal{S}_i$, where $\mathcal{S}_i$ denotes the set of admissible states for agent $i$. The collective state of all $n$ agents is defined as the joint set $\mathcal{S} = \mathcal{S}^1 \times \cdots \times \mathcal{S}^n$. Given that the vehicles are underactuated, the control input for agent $i$ is denoted as $\boldsymbol{u}_i = [u^i, r^i]^\top \in \mathcal{U}^i$, where the linear velocity is constrained by $|u^i| < u_{\max}$ and the angular velocity by $|r^i| < r_{\max}$. Here, $\mathcal{U}^i$ represents the set of feasible control inputs for agent $i$, and the joint action space for all $n$ agents is given by $\mathcal{U} = \mathcal{U}^1 \times \cdots \times \mathcal{U}^n$. The goal of this work is to develop an optimization algorithm that identifies the optimal control input within $\mathcal{U}$ to enhance cooperative tracking performance. To tackle this optimization challenge, we utilize an MPC scheme. The latter foresees the prediction of the evolution of the systems involved in time; therefore, we move to a time-discrete formulation for the optimization problem.

Assuming that at a generic time instant $t_k$ agent $i$ triggers the optimization with *belief state* represented by $\boldsymbol{b}_k^i = (\hat{\mathcal{S}}_k^i, \hat{\boldsymbol{\xi}}_k^i, \boldsymbol{P}_k^i)$ which encompasses both the state of the multi-agent system and the state of the target estimated by agent $i$. Our objective is to select actions over time to minimize the expected cost. Specifically, the MPC scheme attempts to solve an optimal control problem at every time instant $k$ using the information from the identified parameters over a finite future horizon of $H$ steps. Suppose each agent has a control vector of the following form that needs to be optimized

$$\bar{U}_{i,k} = [\boldsymbol{u}_k^i, \boldsymbol{u}_{k+1}^i, \ldots, \boldsymbol{u}_{k+H-1}^i]^\top \in \mathcal{U}^i. \tag{27}$$

To assign a cost to the control actions, we define a generic objective function for the agent $i$ as

$$J_k^i(\bar{U}_{i,k}) = \sum_{h=k}^{k+H-1} f_h(\hat{\boldsymbol{b}}_h^i, \boldsymbol{u}_h^i) + f_H(\hat{\boldsymbol{b}}_{k+H}^i) \tag{28}$$

where $f_h(\cdot)$ represents the transition costs at any time instant and $f_H(\cdot)$ represent the terminal cost. Finally, the optimal control sequence is obtained as

$$\bar{U}_{i,k}^* = \underset{\bar{U}_{i,k}\in\mathcal{U}_i}{\arg\min}\, J_{i,k}(\bar{U}_{i,k}), \tag{29}$$

but only the first vector $\boldsymbol{u}_k^i$ of the control sequence is applied. This process is repeated at the next time step, $k+1$, by incorporating the updated parameter estimates and solving the optimization problem again, with the moving horizon shifted forward by one step. The key advantage of this repeated online optimization lies in the feedback it provides. This approach, known as receding horizon control, is appealing as it offers lookahead capability without the technical complexities associated with infinite horizon control. Nevertheless, the term $f_h(\hat{\boldsymbol{b}}_{i,h}, \boldsymbol{u}_{i,h})|_{h=k:k+H-1}$ is often difficult to obtain, particularly due to the large belief-state space. Consequently, approximation methods are necessary.

A common method for handling uncertainty in target estimation is Nominal Belief Optimization (NBO), which propagates the estimated state using an assumed model. While computationally efficient, NBO can lead to overconfident solutions when the target dynamics are uncertain, as in our case. Since Monte Carlo methods are computationally prohibitive for multi-agent underwater systems with limited resources (Robert et al., 1999), we adopt the UT (Julier and Uhlmann, 2004) to capture the uncertainty in the target state. At each time step $k$, agent $i$ applies the UT to its current target state belief, represented by the mean $\hat{\boldsymbol{\xi}}_k^i$ and covariance $\mathbf{P}_k^i$, generating a set of $L = 2n_\xi + 1$ sigma points. These points represent plausible realizations of the target's future state and are propagated forward using the motion model in (3) over the prediction horizon $H$. For each resulting sigma-point trajectory, the agent evaluates a cost function $f_h$ at every time step $h$, which depends on the sigma-point-based belief ${}^{(l)}\hat{\boldsymbol{b}}_h^i$ and the candidate control input $\boldsymbol{u}_h^i$. The cumulative cost ${}^{(L)}\boldsymbol{J}$ is computed considering $L$ possible realizations of the target state, according to the candidate control sequence $\bar{U}_{i,k}$, such that

$${}^{(L)}\boldsymbol{J}_k^i(\bar{U}_{i,k}) = \sum_{h=k}^{k+H-1}\sum_{l=1}^{L}[\omega_l f_h({}^{(l)}\hat{\boldsymbol{b}}_h^i, \bar{\boldsymbol{u}}_h^i)] + f_H(\hat{\boldsymbol{b}}_{k+H}^i), \tag{30}$$

where $\omega_l$ is the UT weight associated with sigma point $l$ . The optimal control sequence is obtained by solving the following optimization problem:

$$\bar{U}_{i,k}^* = \underset{\bar{U}_{i,k}\in\mathcal{U}_i}{\arg\min}\, {}^{(L)}\boldsymbol{J}_k^i(\bar{U}_{i,k}). \tag{31}$$

The sigma points are generated using the standard UT formulation as

$$\xi^{(0)} = \boldsymbol{A}\xi, \quad \omega^{(0)} \text{ is a tuning parameter}, \tag{32}$$

$$\xi^{\pm(i)} = \xi^{(0)} \pm \sqrt{\frac{n_\xi}{1-\omega^{(0)}}}\,\boldsymbol{k}_j, \quad \omega^{(\pm i)} = \frac{1-\omega^{(0)}}{2n_\xi}, \tag{33}$$

where $\boldsymbol{k}_j$ denotes the $j$-th column of the matrix $\boldsymbol{K} = \mathbf{P}_{i,k}\mathbf{P}_{i,k}^\top$, obtained via Cholesky decomposition. The matrix $\boldsymbol{A}$ maps the sigma points into the belief space used for trajectory generation. For clarity, we omit the superscript ${}^{(L)}$ in the cost function notation in the remainder of the text, but the reader should keep in mind that each cost is evaluated across $L$ sigma-point-based belief trajectories.

#### 2.4.1. Cost functions definition

As introduced in Section 1.1, we implemented an MOO. The rationale for the choice of the cost functions hinges on two key facts: to ensure observability and accurate identification of the target state (particularly given the presence of significant noise and potential outliers), we need to move the agents such that the regressor in (12) needs to be *well-conditioned*. Additionally, to enhance the estimation process while continuously tracking the target, agents must minimize the impact of noise, which depends on the range from the target as outlined in (10).

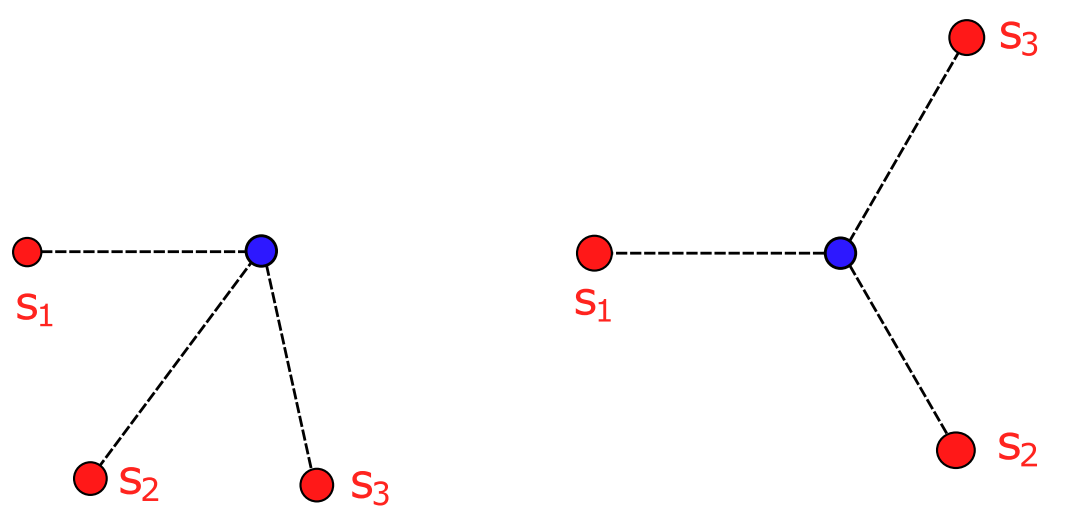

**Fig. 3. Optimal geometric configurations:** The figure shows the eight possible optimal solutions in the case of three sensors corresponding to two geometric configurations. The other optimization outputs and the other infinite solutions can be obtained by rigidly rotating the configuration of an arbitrary angle. Notice that the labels are interchangeable and that the sensors can be placed everywhere in the ray they belong to.

To optimize the position of the agents amounts to maximizing the least singular value $\sigma_1$ of the following matrix

$$\mathbf{\Phi} = \begin{bmatrix} \sin(\beta_i) & -\cos(\beta_i) \\ \sin(\beta_j) & -\cos(\beta_j) \\ \dots & \dots \\ \sin(\beta_n) & -\cos(\beta_n) \end{bmatrix} \in \mathbb{R}^{2\times(n)}. \tag{34}$$

Is it worth noticing that the least singular value of the regressor is $\sigma_1(\mathbf{\Phi}) = \frac{1}{\kappa(\mathbf{\Phi})}$ where $\kappa(\mathbf{\Phi})$ is the *conditioning* of the matrix $\mathbf{\Phi}$ and it may be used as a measure of how far $\mathbf{\Phi}$ is from being singular.

Due to space limitations, we do not elaborate further, as this topic has been extensively studied in Moreno-Salinas et al. (2012), Dogancay (2022), Moreno-Salinas et al. (2016) using the determinant of the FIM as an optimization index. It is important to point out that our proposed index closely aligns with this approach: maximizing the least singular value of $\mathbf{\Phi}$ (or equivalently minimizing its conditioning) indirectly improves the determinant of the FIM since it is the product of the singular values of $\mathbf{\Phi}$. For the sake of clarity, we just provide in Fig. 3 the optimal sensors-target configurations considering three sensors. Since the optimization is based on a look-ahead optimization scheme, to construct the objective function we are interested in the expected value of the matrix $\hat{\mathbf{\Phi}}(\bar{\boldsymbol{p}}_s)$ over the horizon $H$, which depends upon the position of all the agents such that $\bar{\boldsymbol{p}}_s = [\boldsymbol{p}_{s,i}]_{i=1:n}$. Given that the optimization is cast in the form of a minimization problem, the first loss function can be defined as

$$^{(g)}J^i_k = \frac{1}{\sigma_1(\hat{\mathbf{\Phi}}_i(\bar{\boldsymbol{p}}_s)))}. \tag{35}$$

The superscript $(g)$ identifies that the objective function is related to optimizing the geometry of the agents.

On the other hand, to reduce the impact of noise the agents should reduce the distance to the target. To this extent, the second objective function is defined as

$$^{(d)}J^i_{i,k} = \frac{\hat{d}_i}{d^*} \quad s.t. \quad \hat{d}_i = \|\hat{\mathbf{p}} - \mathbf{p}_{s,i}\|. \tag{36}$$

where $d^*$ is the desired range to the target, since, in monitoring scenarios, agents cannot get too close to the target. The superscript $(d)$ identifies that the objective function is related to optimizing the distance to the target. Lastly, in multi-agent settings, it is fundamental to consider that the connectivity of the graph is maintained and possibly improved. To quantify the connectivity of the graph, we use the second smallest eigenvalue of the Laplacian matrix $L$ defined in (26) as a performance index, leading to the following cost function definition

$$^{(c)}J^i_{i,k} = \sum_{j=1}^{N_i} \frac{1}{\sigma_2(L(\hat{\rho_{ij}}))}, \tag{37}$$

the superscript $(c)$ identifies that the cost function is related to the connectivity of the graph.

*Remark:* In the prediction phase, we use the acoustic model from Section 2.3 to estimate the expected SNR $\hat{\rho}_{ij}$ between nodes for objective evaluation. While more complex models could be applied, selecting the most accurate one is beyond the scope of this work. We assume the availability of a reasonable acoustic propagation model suitable for the marine environment considered. The current acoustic propagation model is valid for deep water environments. Extension to shallow water requires the incorporation of multipath propagation models, which is beyond the scope of this work but represents an important direction for future research.

Finally, the cumulative cost function over the horizon, considering also the terminal cost, is computed as

$$\begin{aligned} J^i_k(\bar{U}_{i,k}) = \textstyle\sum_{h=k}^{k+h-1} [\alpha^{(g)} J_{i,h} + (1-\alpha)^{(d)} J_{i,h} + \gamma^{(c)} J_{i,h}] + \dots \\ \dots + (\|\boldsymbol{p}_H - \boldsymbol{p}_{s,H}\| - d^*) \end{aligned} \tag{38}$$

where the trade-off weight $\alpha \in [0, 1]$ balances between accurate target estimation and noise resilience, while $\gamma \in [0, 1]$ is a regularization terms that penalizes constraint violations. More insight about this specific implementation of the MOO can be seen in Tiranti et al. (2024), where the relationship between the cost functions is deeply analyzed.

*Remark:* Notice that the weight $\gamma$ acts as a regularization term, used to weight different constraint functions. In this setup, the cost function in (37) plays the role of a *soft constraint* on the optimization problem (Babar and Baglietto, 2021). The tuning depends upon the expected acoustic modem performances according to the size and the morphology of the area to be monitored. Assuming that the communication is possible without any particular network topology, $\gamma$ can be low, otherwise, a high value of $\gamma$ will force the agents to maintain the topology of the network assigned in advance.

#### 2.4.2. Sequential multi-agent decision making

The monitoring task is approached as a cooperative game where agents pursue both individual and collective objectives. Due the uncertainties introduced by the underwater environment and the communication constraints, centralized and purely decentralized methods are impractical. To overcome this, we employ a fully distributed strategy based on sequential multi-agent decision-making (Gmytrasiewicz and Doshi, 2005), where each agent optimizes and shares its control sequence, as depicted in Fig. 4. This promotes cooperation, aligns the understanding of the environment among agents, and ensures robustness to packet loss by enabling behavior prediction during communication losses.

Let us define $\bar{U}_i$ with $i \in [1, n]$ as the single agent control sequence based on the *belief state* $t_k$ that is

$$\bar{U}_i = (\boldsymbol{u}^i_k, \boldsymbol{u}^i_{k+1}, \dots, \boldsymbol{u}^i_{k+H-1}). \tag{39}$$

We will omit when possible the subscript $k$ for the sake of clarity, but notice that everything depends on the discrete time $t_k$ considered in the optimization process.

To illustrate how multi-agent planning would ideally work, we consider a decentralized MPC formulation where the optimization starts with the first agent computing its optimal control sequence $U^*_1$ as

$$\bar{U}^*_1 = \arg\min J^1(\boldsymbol{s}^1, \hat{\boldsymbol{\xi}}^1, \bar{U}^*_2, \dots, \bar{U}^*_n). \tag{40}$$

Here, $J^1$ denotes the cumulative cost introduced in (38), which depends on the state of the agent, the estimated target state, and the predicted behaviors of neighboring agents, assumed to follow their respective optimal sequences $\bar{U}^*_j$. This formulation is theoretically consistent with decentralized MPC, where each agent optimizes its policy under the assumption that its neighbors will execute optimal plans.

However, applying this approach in our underwater scenario presents two critical challenges. First, the high uncertainty introduced by the underwater environment makes it difficult to maintain accurate and up-to-date knowledge of neighbors' states. Second, the TDMA-based communication protocol inherently introduces synchronization delays and limits real-time data exchange. As a consequence, assuming that each agent knows the exact control sequence of its neighbors during execution becomes impractical. To address these limitations, we adopt a recursive distributed implementation of the control framework. In this approach, agents explicitly communicate their predicted control sequences to neighbors, allowing for iterative coordination and local replanning. This strategy enables a more robust and realistic cooperation mechanism under constrained communication, making it suitable for real-world underwater deployment where centralized or fully synchronized decentralized approaches fail to scale or adapt.

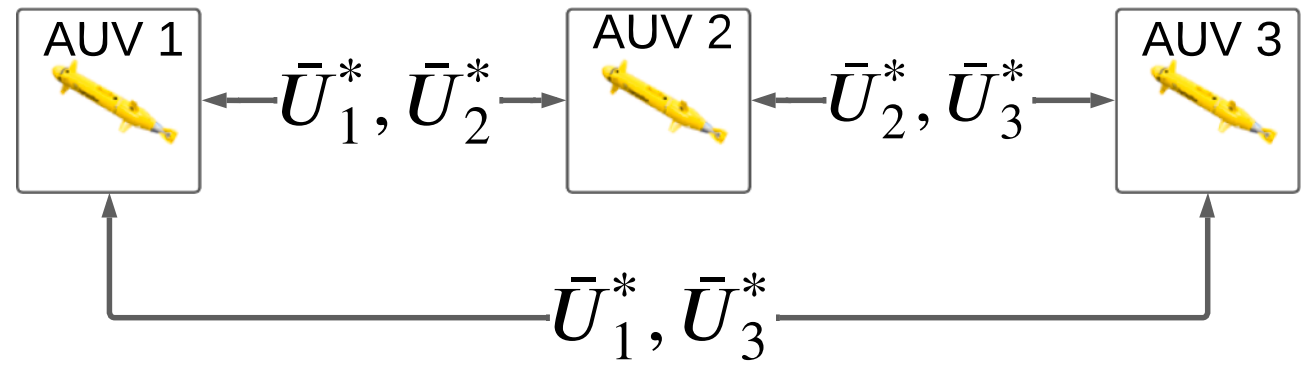

**Fig. 4. Sequential multi-agent decision making:** The image illustrates a block diagram that describes the sequential multi-agent decision-making framework considering a case with three autonomous agents with omnidirectional acoustic modems. The symbol $\bar{U}_i^*$ refers to the optimal control sequence computed by agent $i$, while $\bar{U}_i^R$ refers to the policy of intent of agent $i$.

In sequential planning frameworks, the control sequences for a sequence of agents from $i$ to $i+j$ is a *decision epoch* defined as $\bar{U}_{i:i+j} = (\bar{U}_i, \bar{U}_{i+1}, \dots, \bar{U}_{i+j})$. To implement such a strategy, we need to tackle two main problems: for each decision epoch, each agent knows only the optimal control sequence computed by the previous agents in the planning sequence; we might lose some of the control sequences exchanged due to packet loss. We propose to use previously exchanged sequences to approximate the intention of the agents when up-to-date information is not available. In this context, when agent $i$ triggers the optimization at time $t_k$, $\forall$ agent $j \neq i \in N_i(t)$, we have a remaining sequence of control actions computed at the previous optimization time $t_{k-1}$ such that

$$\bar{U}^j_{k-1} = (\boldsymbol{u}^j_{k-1}, \boldsymbol{u}^j_k, \dots, \boldsymbol{u}^j_{k+H-2}). \tag{41}$$

This quantity provides insight about agent $j$ future intentions, thus can be used at time $k$ to inform agents $1 : i | i < j$. From an algorithmic point of view we need to remove the old control command applied in $t_{k-1}$, then the length of the sequence became equal to $(H-1)$; and thus we add the control action $\boldsymbol{u}_{j,k+H}$ as a heuristic single agent base control $\bar{\boldsymbol{u}}_{j,e}$ for obtaining the policy of intent as

$$\bar{U}^R_j = (\boldsymbol{u}^j_k, \dots, \boldsymbol{u}^j_{k+H-2}, \boldsymbol{u}^j_e). \tag{42}$$

The generalized form of the sequential multi-agent planning algorithm can be written as

$$\bar{U}_i^* = \arg\min J_i(\boldsymbol{b}_k, \bar{U}_1^*, \dots, \bar{U}_{i-1}^*, \bar{U}_i, \bar{U}_{i+1}^R, \dots, \bar{U}_n^R). \tag{43}$$

To implement this procedure, the agents must exchange a message with the following format $[\boldsymbol{s}^i_k, \boldsymbol{u}^i_k, \dots, \boldsymbol{u}^i_{k+H-1}]$.

*Remark:* The computational complexity for one agent optimization is $\mathcal{O}(|u^H|)$ and using the proposed approach the complexity for $n$ agents becomes $\mathcal{O}(|u^H| \cdot n)$, way less than the complexity $\mathcal{O}(|u^H|^n)$ of a *greedy search* over a solution space for $n$ agents (Fig. 4) .

### 2.5. Implementing the optimization problem for real-case scenarios

Two main aspects must be considered while porting the proposed distributed control framework in a multi-AUV system: the length of the optimization time step of the optimization problem, the control choices available to each agent.

We start by defining the optimization time step, $T_o$, which must be tuned based on the communication protocol, in this case, TDMA. Since each agent is assigned a time slot $T_s$, a full communication round requires $T_f = nT_s$. However, due to the iterative nature of the distributed optimization and the presence of packet loss, a single round is typically not enough for agents to reach a consistent planning strategy for cooperative target tracking. To ensure that sufficient information is exchanged before computing a new control sequence, the optimization time step is defined as:

$$T_o = (nT_s) \cdot n^{\text{iter}}, \tag{44}$$

where $n^{\text{iter}}$ is the number of communication rounds needed for the algorithm to converge. This design ensures that the system has time to gather the necessary data while balancing responsiveness and coordination performance. The only necessary condition to is the following:

$$T_o >> T_s \tag{45}$$

The optimization problem for a single agent has a computational complexity of $\mathcal{O}(|\mathcal{U}_i|^H)$, where $|\mathcal{U}_i|$ denotes the cardinality of the control set for the agent $i$. Follows that the size of the control set must be limited to have a tractable optimization problem. In our scenario, we want to optimize both the heading of the vehicle and its surge velocity. For the latter, we consider three possibilities, such that the relative set is defined as

$$\mathcal{U}^i_u = \{0, u^i_e, u_{max}\}. \tag{46}$$

Notice that $u^i_e$ is computed heuristically according to the range from the target

$$u^i_e = \frac{|d^* - \hat{d}_i|}{T_o} + |\hat{\boldsymbol{v}}|, \tag{47}$$

where $\hat{\boldsymbol{v}}$ is the estimated target velocity. In this way, we have a limited control set but which can adapt to changes in the target motion. For the heading control, we specify a heading change range $[-\Delta\theta_{\max}, +\Delta\theta_{\max}]$ and discretize it into $n_\theta$ steps, obtaining the following control set

$$\mathcal{U}^i_\theta = \{\Delta\theta_1, \dots, \Delta\theta_{n_\theta}\}, \tag{48}$$

which leads to the following definition for the overall control set

$$\mathcal{U}^i = \mathcal{U}^i_\theta \times \mathcal{U}^i_u. \tag{49}$$

The resulting combinatorial problem is solved using a nested branch and bound (BnB) implementation, because knowing the nature of the problem, we can further simplify the optimization using specific *pruning strategies*. The resulting control decisions can be converted to waypoints and fitted with polynomial trajectories suitable for AUV path-tracking algorithms.

Finally, we present the complete optimization problem for multi-agent guidance in cooperative target tracking scenarios, considering a generic agent $i$. In the following formulation, we explicitly state all the constraints to the optimization problem according to the proposed formulation, also incorporating additional real-world constraints and safety constraints:

$$\begin{aligned}
&\bar{U}^*_{i,k}(\boldsymbol{b}_{i,k}) = \dots \\
&\dots = \underset{\bar{U}_{i,k} \in \mathcal{U}^i}{\arg\min} \quad J_i(\bar{\boldsymbol{p}}_s, \hat{\boldsymbol{\xi}}^i, \boldsymbol{P}^i, \bar{U}_1^*, \dots, \bar{U}_{i-1}^*, \bar{U}_i, \bar{U}_{i+1}^R, \dots, \bar{U}_n^R). \\
&\qquad \text{s.t.} \quad \bar{\boldsymbol{p}}_s = [\boldsymbol{p}^i_s]_{i=1:n}, \\
&\qquad\qquad \begin{bmatrix} \dot{\boldsymbol{p}}^i_s \\ \dot{\theta}^i_s \end{bmatrix} = \boldsymbol{\Theta}(\theta_s) \begin{bmatrix} u^i \\ 0 \\ r^i \end{bmatrix}, \\
&\qquad\qquad |u^i| \leq u_{max}, \\
&\qquad\qquad |r^i| \leq r_{max}, \\
&\qquad\qquad \hat{\boldsymbol{\xi}}^i_{k+1} = \boldsymbol{F} \hat{\boldsymbol{\xi}}^i_k, \\
&\qquad\qquad \boldsymbol{P}^i_{k+1} = \boldsymbol{F} \boldsymbol{P}^i_k \boldsymbol{F}^\top, \\
&\qquad\qquad d^* < d_i < d_M, \\
&\qquad\qquad d_{ij} < d_S, \forall j \in N_i
\end{aligned} \tag{50}$$

The first constraints are related to agents' kinematics, including maximum linear and angular velocity constraints. Then follows the constraints related to the dynamics of the target and its estimate. Contrary to the model in (3), we switched to a discrete time formulation in which $\boldsymbol{F}$ is an appropriate approximation of the matrix $\boldsymbol{A}$. Finally, $d_M$ is the maximum acceptable distance from the target (bounded by acoustic signal attenuation limits in PAM applications), and $d_S$ is the minimum safety distance between agents.

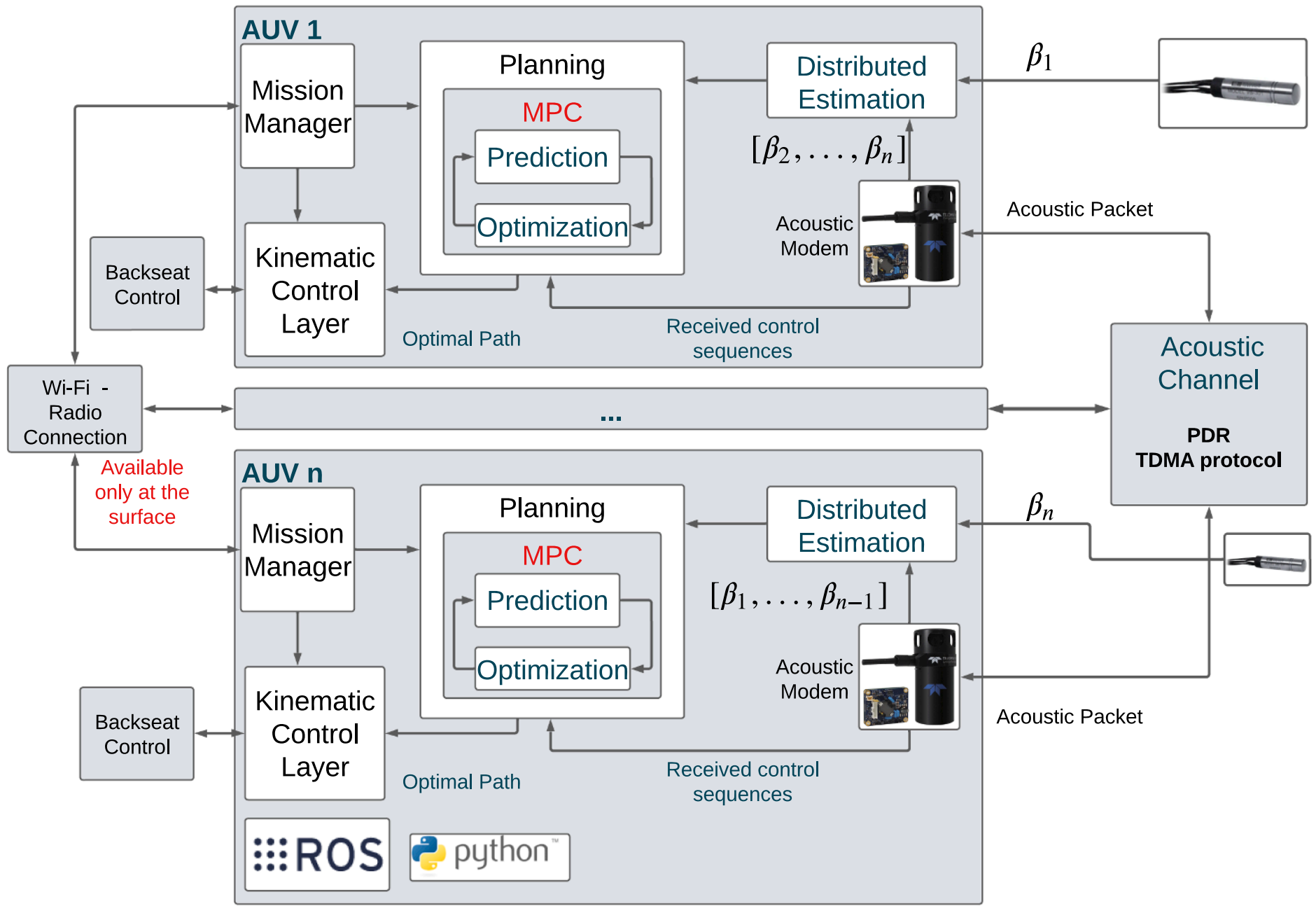

**Fig. 5. Distributed software architecture:** The image illustrates the software architecture developed for the proposed distributed framework. In this work, the focus is on the MPC block, which acts as a high-level planning module within the guidance layer. The planning module receives assignments from the mission manager and incorporates information from the tracking module and the acoustic modem to generate motion plans over a receding horizon. Each new guidance solution is then passed to the kinematic control layer, which functions as the low-level controller responsible for path tracking through the backseat control. This separation highlights the role of MPC as a planner for cooperative guidance, while conventional controllers ensure trajectory execution. Although validation is conducted in simulation, the software is designed to function with a real multi-agent system. In fact, we included the nodes related to communication (both Wi-Fi and acoustic). Currently, these blocks are replaced with their simulated counterparts; for example, the acoustic channel node is used to model latencies and packet loss based on user-defined parameters. This setup allows us to analyze system performance under varying conditions.

**Table 1**
**Validation - Section 3: Simulation parameters.** Symbols, meanings, values, and units of the simulation parameters used in our case study. Each parameter is reported with three values, corresponding to the three validation scenarios presented in Sections 3.1–3.3.

| Param | Meaning | Sim 1 | Sim 2 | Sim 3 | Unit |
|---|---|---|---|---|---|
| $n$ | Num. agents | 3 | 3 | 3 | - |
| $T_m$ | Meas. sampling period | 1 | 2 | 2 | $s$ |
| $T_s$ | Time slot | 6 | 14 | 14 | $s$ |
| $\sigma_m$ | Meas. uncert. | ±3.5 | ±4.5 | ±6.5 | deg |
| $NL$ | Noise level | 20 | 40 | 50 | dB |
| $PDR$ | Packet deliv. ratio | 95 % | 65 % | 65 % | - |
| $d_S$ | Safety dist. | 25 | 100 | 100 | m |
| $d_M$ | Max dist. from target | 0.5 | 1 | 2 | km |
| $d^*$ | Desired dist. from target | 30 | 500 | 500 | m |
| $u_{max}$ | Max speed | 1.0 | 2.0 | 2.0 | m/s |
| $r_{max}$ | Max turn rate | 0.05 | 0.05 | 0.05 | rad/s |
| $U_0$ | Control set size | 7 | 9 | 9 | - |
| $H$ | Planning Horizon | 4 | 5 | 5 | - |
| $L$ | UT parameter | 2 | 4 | 4 | - |
| $T_o$ | Optimization time step | 60 | 130 | 130 | s |

## 3. Results

In this section, we analyze the performance of the proposed motion planning strategy for cooperative underwater acoustic source tracking and pursuit. The distributed control framework is implemented using the Robotic Operating System (ROS) Noetic version, with Python and C++, on Ubuntu 20.04.5 LTS. The system runs on a computer configured with an Intel® Core™i7-8550U CPU @ 1.80GHz × 8. The overall software architecture is summarized in Fig. 5. It is worth recalling that our simulation omits low-level dynamics. Nevertheless, the guidance performance evaluation remains valid since the path-following controllers typically achieve tracking errors of 1–2 m, which is small compared to our target tracking distances (300 m - 2 km range) .

We solve the optimization problem using a Branch and Bound (BnB) algorithm implemented in Python via the PyBnB library. For the rest of the section, we consider three AUVs equipped with an omnidirectional ($DI = 0$) medium-range acoustic modem with source level $SL = 186$ (dB) and acoustic center frequency $f = 10$ (kHz). To have a realistic simulation, we consider a limited capacity of the communication channel by setting the bit rate at 120 (bps). All validations are conducted considering the access to the acoustic channel governed by the TDMA protocol. Therefore, each agent shares the same notion of time and has a time slot $T_f$ to transmit.

This section is organized as follows. In Section 3.1, as a preliminary validation, we will show three possible operational scenarios in which the proposed methodology can be applied and examine how the weights of the objective functions in (50) influence the outcome of the mission and their importance during sea trials. In Section 3.2, we validate the proposal by considering a realistic and challenging scenario of underwater cooperative target tracking. The goal is to demonstrate that the proposal is robust against uncertainties, partial knowledge of the environment, and potential node failures. Finally, Section 3.3 presents a comparative analysis in which the proposed approach is evaluated against standard benchmarks commonly used in multi-AUV planning and guidance. All simulation parameters are listed in Table 1, with the

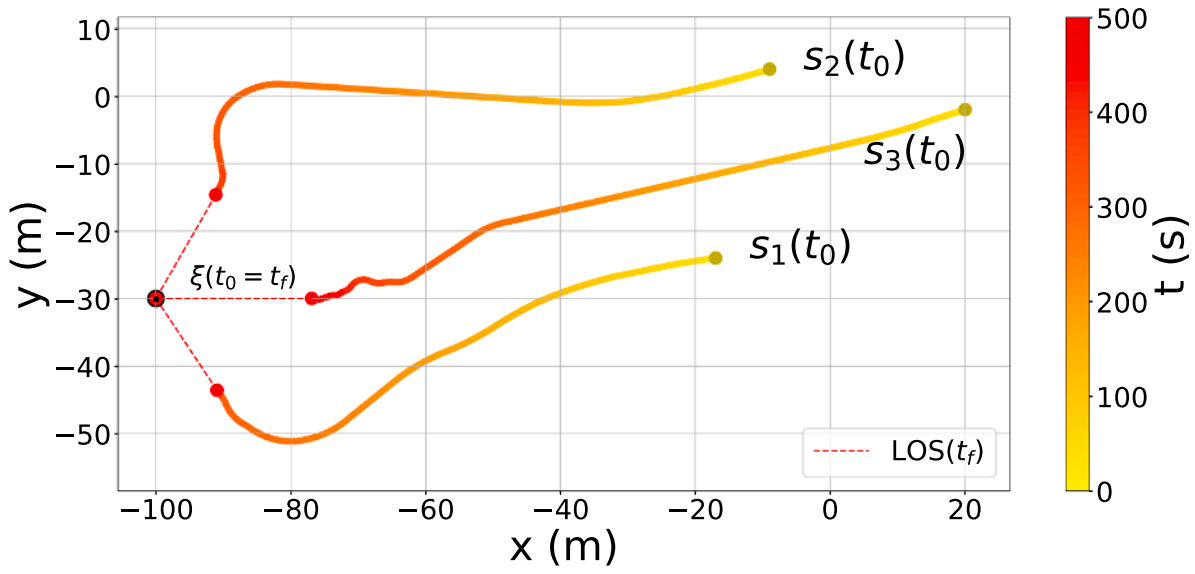

(a) **Scenario 1:** Agent motion for optimal sensor placement ($\alpha = 0.8$, $\gamma = 0.3$). Final configuration matches one of the optima in Fig. 3.

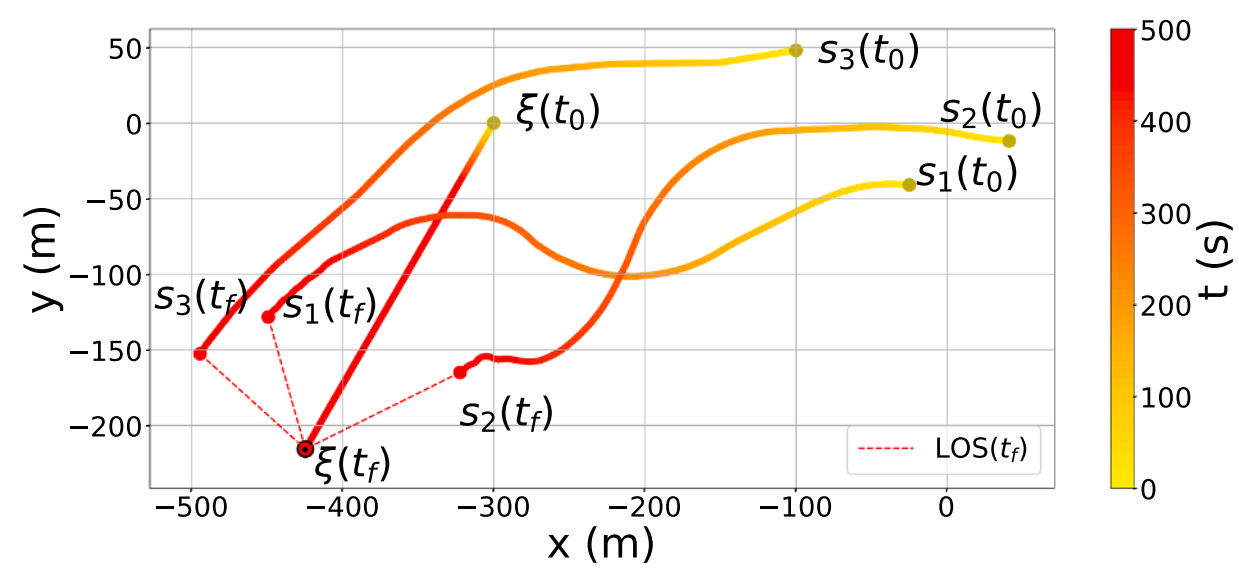

(b) **Scenario 2:** Agent motion for tracking a moving source ($\alpha = 0.5$, $\gamma = 0.3$). See Fig. 6d for objective trend.

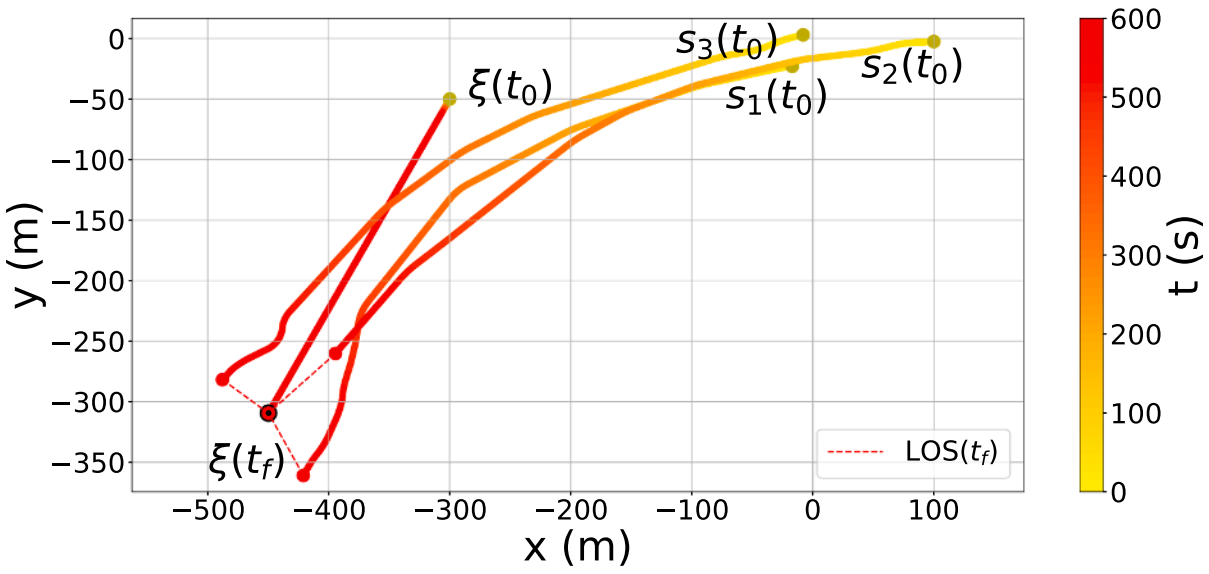

(c) **Scenario 3:** Same tracking setup as (b), but with connectivity preservation. See Fig. 6d for objective trend.

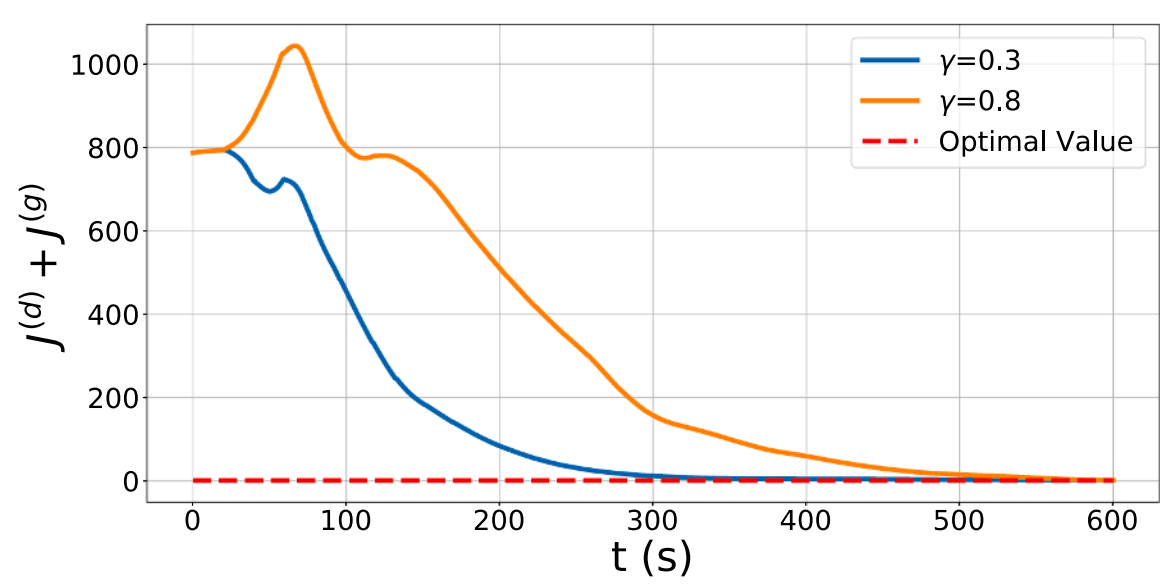

(d) Comparison of $J^{(g)} + J^{(d)}$ for different $\gamma$ values. Lower $\gamma$ leads to faster convergence.

**Fig. 6. Results Section 3.1:** Summary of Scenarios 1–3. Subfigures (a)-(c) show agent motions under different objectives. Subfigure (d) compares convergence of the cumulative cost function under different $\gamma$.

values used in Sections 3.1–3.3 corresponding to Sim 1, Sim 2, and Sim 3, respectively.

### 3.1. Preliminary results

The validation presented in this section serves as an analysis of the effect of the multi-objective optimization on the agents' behavior during the acoustic monitoring mission. Given the applications mentioned in Section 1, we consider two scenarios: optimal sensor placement considering a fixed target (Moreno-Salinas et al., 2012), and tracking and pursuit of a moving target (Nguyen et al., 2021). The key idea of the proposed method is that the tuning of the weights in (50) depends on the specific scenario in which the autonomous agents are employed. Consequently, we will present three different scenarios of underwater acoustic monitoring in which the tuning of the mentioned weights plays a fundamental role in improving the outcome of the mission or inducing a desired behavior. The values of the simulation parameters used are listed in the column *Sim 1* of Table 1.

**SCENARIO 1:** Consider three agents starting from the following positions $\boldsymbol{p}_{s,1} = (-17, -23)^\top$, $\boldsymbol{p}_{s,2} = (-10, 5)^\top$ and $\boldsymbol{p}_{s,3} = (20, -3)^\top$, and a fixed target located at $\boldsymbol{p} = (-30, -100)^\top$. As explained in Section 2, the choice of $\alpha$ is dependent on whether the state of the target needs to be identified or the impact of measurement noise reduced. Assuming that the agents are employed for optimal sensor placement, then the main objective is to minimize the cost function in (35), and this can be easily achieved by letting $\alpha \to 1$. Fig. 6a shows the behavior of the agents for $\alpha = 0.8$ and $\gamma = 0.3$.

**SCENARIO 2:** Consider three agents starting from the following positions $\boldsymbol{p}_{s,1} = (-17, -23)^\top$, $\boldsymbol{p}_{s,2} = (-10, 5)^\top$ and $\boldsymbol{p}_{s,3} = (20, -3)^\top$, and a moving target with starting position at $\boldsymbol{p} = (-30, -100)^\top$. To track a moving target, agents must move in a coordinated manner and stay within a certain vicinity of the target to reduce the impact of measurement noise. This can easily be achieved by decreasing $\alpha$. Assuming that the target moves at a constant velocity with heading $\pi/4$, the behavior of the agents given $\alpha = 0.5$ and $\gamma = 0.3$ is shown in Fig. 6b.

**SCENARIO 3:** Notice that for both cases we used low values of $\gamma$, hence, the agents are free to move while fulfilling the two main objectives. If we increase $\gamma$ the agents will prioritize connectivity preservation. To verify this concept, we consider the scenario with the moving target presented before, but now setting $\gamma = 0.8$. The outcome of the simulator is shown in Fig. 6c.

### 3.2. Challenging scenario

In this section, we consider a simulation scenario that incorporates more realistic challenges to demonstrate the capabilities of the proposed methodology despite the significant challenges posed by the marine environment. This simulation must be viewed as a representative scenario to understand the effect of the proposed planning strategy. To this extent, the target now moves with a trajectory defined by $\boldsymbol{p}(t) = [a \cdot \sin(\omega t + \pi),\ v_n t]^\top$ with a random initial heading. The agents' initial positions are randomly selected given an area of $200m^2$ (the only constraint is the safety minimum distance $d_S$), and the target initial position is $\boldsymbol{p} = (-450, 105)^\top$. In contrast to the previous validations, we now increase packet loss by setting the Packet Delivery Ratio (PDR) to 65 %. To stress the system, after $t = 750s$ we simulate the failure of one of the AUVs. The values of the simulation parameters used are listed in the column *Sim 2* of Table 1. The results are shown in Fig. 7, specifically 7a shows the motion of the agents and the target, Fig. 7b depicts the trend of the tracking error for each AUV and 7c outline the trend of the cu-

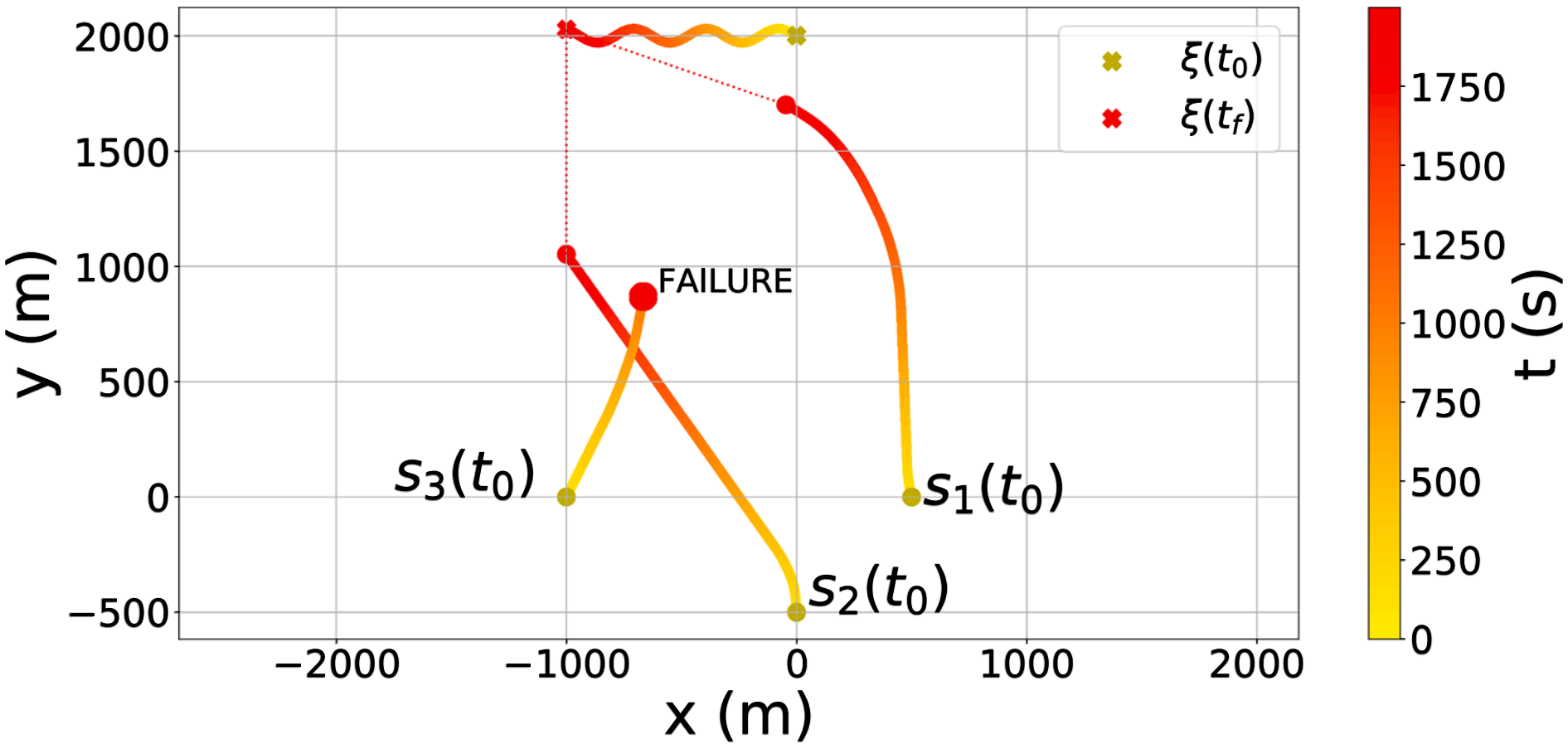

(a) Agent motion during cooperative tracking and pursuit.

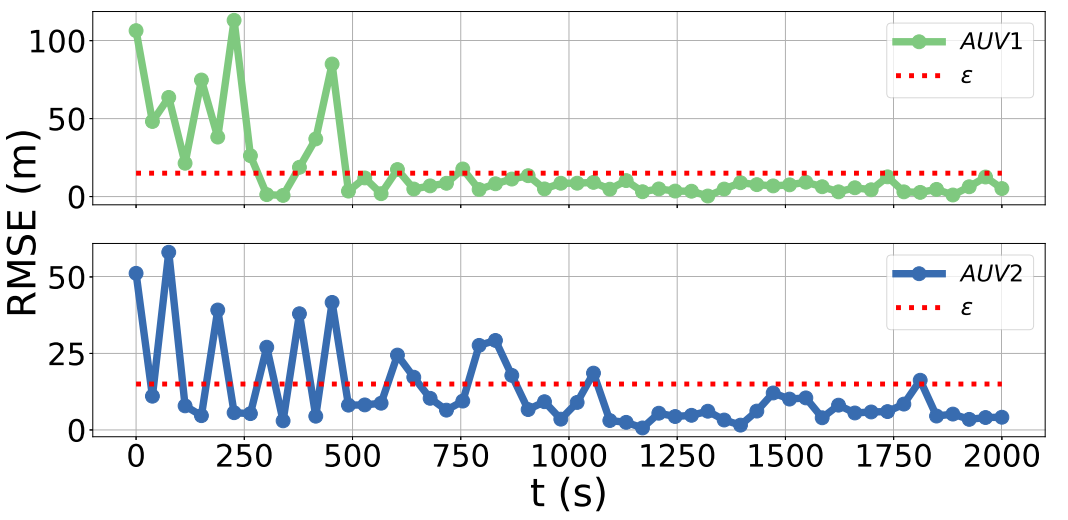

(b) Tracking error for each AUV. The red line is a 15 m error threshold.

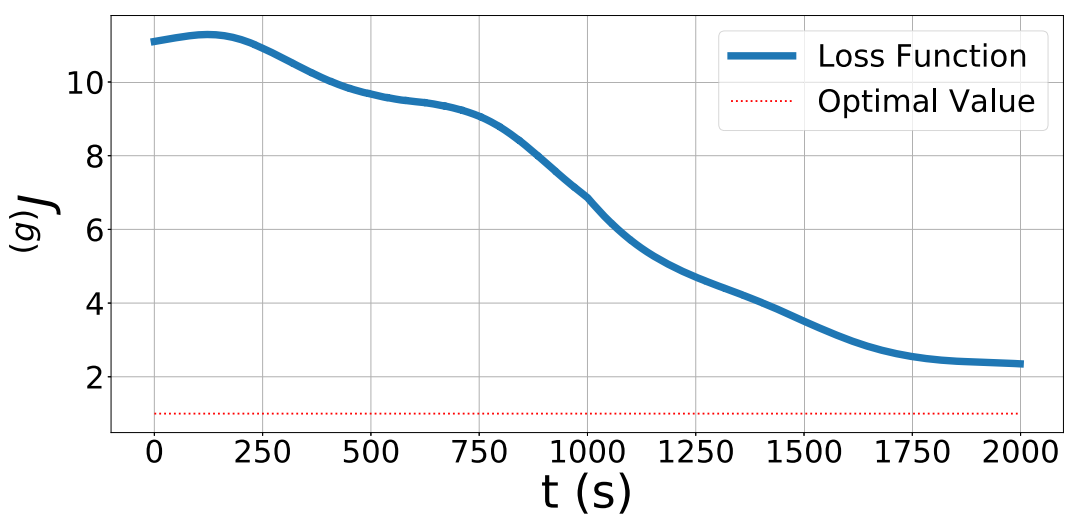

(c) Trend of the cumulative loss function based on sensors-target geometry.

**Fig. 7. Validation Section 3.2 - Performance Analysis, Scenario 2.1.** (a) Motion of the agents and target. (b) Time evolution of tracking error for each AUV. The red dotted line represents a 15 m acceptable error bound under realistic conditions (high noise, failure). (c) Cumulative loss function trend. After an agent failure, the loss initially struggles to converge due to a change in network topology, but the method reconfigures the team toward optimal performance.

mulative loss function. Notice that the convergence of the loss function to zero is not degraded after the failure of one of the AUVs. The weight $\alpha$ is set to 0.5 and $\gamma = 0.25$, ensuring that the agents optimize their geometry while simultaneously pursuing the target, which represents the most challenging case for underwater acoustic monitoring missions. If the reader is interested, an animation video of the simulation can be found at the following link: https://www.youtube.com/watch?v=rd-hykhykRl8

### 3.3. Comparative analysis

To compare the performance of the proposed methodology with established approaches, we consider the following benchmark methods:

**Decentralized MPC (dec-MPC)**: each agent optimizes its trajectory assuming neighboring agents follow an optimal policy, without any exchange of control sequences;

**Distributed Particle Swarm Optimization (DPSO)**: implemented within our distributed control framework, as this technique is commonly employed in cooperative target tracking scenarios (Wang et al., 2023);

**Centralized MPC**: used as a performance baseline, providing a lower bound on tracking accuracy, though it is not feasible in practice for underwater networks with communication constraints.

To compare the proposed benchmarks to our approach, we evaluate system performance across five realistic cooperative target tracking scenarios, using the simulation parameters listed under the *Sim 3* column in Table 1. The agents' initial positions are set as follows: $\boldsymbol{p}_{s,1} = (500, 0)^\top$, $\boldsymbol{p}_{s,2} = (0, -500)^\top$, $\boldsymbol{p}_{s,3} = (-1000, 0)^\top$. To challenge the system's robustness, the target does not follow a straight-line trajectory, but instead moves along one of two time-dependent paths: $\boldsymbol{p}(t) = [a \cdot \sin(\omega t + \pi),\ v_n t]^\top$ or $\boldsymbol{p}(t) = [v_n t,\ a \cdot \sin(\omega t + \pi)]^\top$, where $a$ and $v_n$ are configurable parameters that define the oscillation amplitude and nominal speed, respectively. The initial heading and motion parameters for each of the five scenarios are detailed in Table 2. These were selected to emulate a variety of operational contexts, including variations in environmental noise and target behavior, thus enabling a comprehensive performance comparison across methods. For each scenario and method, we consider the average tracking error and the standard deviation, together with the average optimization time and the average formation quality index. The results are condensed in Table 3, the values are an average of 10 simulations to reduce the impact of random noise on the data.

### 3.4. Comments and dicussion

The proposed distributed motion planning strategy has proven its validity for reaching optimal sensors-target configurations for tracking underwater acoustic sources. The proposal works both if the target is fixed or moving, as shown in the preliminary validation in Section 3.1, since both in SCENARIO 1 and SCENARIO 2 the agents reach an optimal configuration at the end of the simulation. In SCENARIO 3, we demonstrated that the weight $\gamma$ (see Eq. 50) can be used as a regularization term to decide how much the communication constraints affect the motion planning process. Notice that the agents are moving almost in a fixed formation, given the initial positions and the assigned network topology. This ensures better network connectivity at the expense of the convergence to the optimal solution, as can be seen in Fig. 6c, where the convergence of the sum of objective functions in (35) and (36) is shown for the two values of $\gamma$ used in the previous simulations. The choice of $\gamma$ depends on the expected acoustic modem performances given the characteristics of the marine area of interest.

**Table 2**
**Validation Section 3.3 - Simulation parameters:** The table lists the symbols associated with the simulation parameters, their meaning, and their value for each scenario considered. These parameters can be used to define a specific monitoring scenario; in particular, they are used to define the dynamics of the target.

| Parameter | Meaning | Sim 1 | Sim 2 | Sim 3 | Sim 4 | Sim 5 |
|---|---|---|---|---|---|---|
| $NL$ | Ambient noise level (dB) | 30 | 30 | 60 | 50 | 40 |
| $p$ | Target initial position (m) | [0,-350] | [400, 0] | [400, 0] | [-600,25] | [250,350] |
| $v_n$ | Target nominal velocity (m/s) | 0.2 | 0.5 | 0.15 | 0.35 | 0.35 |
| $\theta_0$ | Target initial heading (rad) | $\frac{\pi}{2}$ | $\pi$ | $\pi$ | $-\frac{\pi}{2}$ | $\pi$ |
| $a$ | Amplitude (-) | 0.3 | 0.1 | 0.2 | 0.2 | 0.3 |
| $\omega$ | Angular velocity (rad/s) | 0.03 | 0.01 | 0.01 | 0.02 | 0.03 |

**Table 3**
**Validation Section 3.3 - Performance comparison across control strategies:** The table reports the average tracking error, average optimization time, and Formation Quality Index (FQI) for each control strategy and scenario. The centralized MPC serves as a baseline. Bold values indicate the best performance among non-centralized methods.

| Scenario | Method | Tracking Error [m] | Optimization Time [s] | FQI (-) |
|---|---|---|---|---|
| Sim 1 | **MPC (centralized)** | $7.8 \pm 5.9$ | N/A | N/A |
| | dec-MPC | $11.6 \pm 9.3$ | 16.0 | 10.5 |
| | PSO (distributed) | $10.5 \pm 7.2$ | 7.2 | 10.2 |
| | **PROPOSAL** | **10.4 ± 8.0** | **8.4** | **8.5** |
| Sim 2 | **MPC (centralized)** | $13.2 \pm 10.6$ | N/A | N/A |
| | dec-MPC | $25.0 \pm 22.1$ | 27.7 | 13.3 |
| | PSO (distributed) | $33.7 \pm 24.3$ | **10.7** | 16.7 |
| | **PROPOSAL** | **13.4 ± 12.6** | 13.2 | **11.4** |
| Sim 3 | **MPC (centralized)** | $13.2 \pm 10.6$ | N/A | N/A |
| | dec-MPC | $28.3 \pm 31.2$ | 30.3 | 8.9 |
| | PSO (distributed) | $41.9 \pm 39.3$ | 11.6 | 20.6 |
| | **PROPOSAL** | **14.4 ± 14.2** | **6.68** | **8.4** |
| Sim 4 | **MPC (centralized)** | $15.0 \pm 13.5$ | N/A | N/A |
| | dec-MPC | $24.1 \pm 22.6$ | 31.4 | 12.6 |
| | PSO (distributed) | $59.2 \pm 53.0$ | 13.9 | 19.3 |
| | **PROPOSAL** | **22.9 ± 21.5** | **9.7** | **9.5** |
| Sim 5 | **MPC (centralized)** | $10.5 \pm 5.0$ | N/A | N/A |
| | dec-MPC | $20.5 \pm 17.7$ | 24.7 | 27.3 |
| | PSO (distributed) | $37.9 \pm 32.3$ | 13.7 | 43.0 |
| | **PROPOSAL** | **14.7 ± 14.9** | **8.5** | **25.5** |

In the successive validation, we stressed the system's performance by reproducing a challenging scenario in which the distances involved are higher. Moreover, three unwanted events happen: the PDR is decreased to 65 %, the target does not move at a constant velocity, and AUV 2 fails after $t = 750s$. Despite the difficulties encountered in this case, the agents reach an almost optimal configuration for target tracking.

In Section 3.3, we compared the performance of the proposed methodology against common approaches used in multi-agent planning problems. Our approach consistently outperforms the alternatives across all scenarios, achieving results that are competitive with the centralized implementation. Moreover, the dec-MPC method is likely to face greater difficulties in real-world settings, where navigation inaccuracies can significantly affect performance due to the lack of explicit information exchange between agents.

Finally, let us conclude by commenting on the computational complexity and optimization time. For our implementation with $|\mathcal{U}^i| = 21$ (7 heading × 3 speed choices), H = 5, and n = 3, this represents a reduction from approximately $4 \times 10^{19}$ to $2 \times 10^7$ operations. In fact, as shown in Table 3, our method achieves optimization times of 6.68-13.2 seconds across scenarios, compared to 24.7–31.4 seconds for decentralized MPC, showing that the sequential framework's efficiency depends on agents reusing previously computed control sequences from neighbors, as described in (42) and (43), rather than solving independent optimization problems.

## 4. Conclusions

This manuscript tackled the problem of planning and guidance for a team of autonomous underwater vehicles. Each agent measures the Direction of Arrival (DoA) of the acoustic signal and shares it with its neighbors, adopting a share and estimate paradigm. This approach avoids the need for aggressive maneuvers by relying on geometric diversity among agents to ensure observability, and it eliminates single points of failure through distributed sensing and estimation. To support this strategy, we employ a motion planning scheme based on Distributed Model Predictive Control (DMPC), designed to guide the team toward optimal geometric configurations with respect to the target. The control framework is implemented in a fully distributed, sequential multi-agent decision-making architecture that accounts for the constraints imposed by acoustic communication. The proposed method demonstrated robust performance in realistic simulation scenarios, accounting for acoustic communication constraints, packet loss, and agent failures. It consistently outperformed standard approaches and showed strong adaptability to changes in the target trajectory or network topology. The framework can also be extended to handle multi-target scenarios, for example, by assigning different targets to different agents or by jointly optimizing the team's behavior to track multiple targets simultaneously. Future work will address the associated scalability challenges and focus on integrating advanced estimation and data association techniques to ensure reliable performance under communication constraints. Moreover, the proposal can be extended to three-dimensional missions by including depth and pitch in the state and control formulations. We acknowledge that formal convergence guarantees for our DMPC approach under acoustic communication constraints remain unestablished. While empirical evidence demonstrates robust convergence behavior, proving theoretical convergence for distributed MPC systems with partial observability and communication constraints represents an important direction for future theoretical research.

## CRediT authorship contribution statement

**Andrea Tiranti:** Writing – original draft, Visualization, Validation, Software, Methodology, Investigation, Data curation, Conceptualization; **Francesco Wanderlingh:** Writing – review & editing, Supervision; **Enrico Simetti:** Writing – review & editing, Supervision, Conceptualization; **Marco Baglietto:** Methodology; **Giovanni Indiveri:** Supervision, Project administration, Funding acquisition, Conceptualization; **Antonio Pascoal:** Writing – review & editing, Supervision, Conceptualization.

## Data availability

Data will be made available on request.

## Declaration of competing interest

During the preparation of this work the author(s) used ChatGPT in order to improve language and readability. After using this tool/service,

the author(s) reviewed and edited the content as needed and take(s) full responsibility for the content of the publication.